\documentclass[a4paper,10pt]{article}

\newcommand{\bfa}[1]{\mbox{\boldmath${#1}$}}

\newcommand{\bea}{\begin{eqnarray}}
\newcommand{\eea}{\end{eqnarray}}
\newcommand{\bnn}{\begin{eqnarray*}}
\newcommand{\enn}{\end{eqnarray*}}
\newcommand{\be}{\begin{equation}}
\newcommand{\ee}{\end{equation}}

\newcommand{\nn}{\nonumber}

\def\PACS{\par\leavevmode\hbox {\it PACS:\ }}%
\def\MSC{\par\leavevmode\hbox {\it MSC:\ }}%
\def\UK{\par\leavevmode\hbox {\it Keywords:\ }}%

\begin{document}

\title{The Poincar\'e mass operator in terms of a hyperbolic algebra}

\author{S. Ulrych\\Wehrenbachhalde 35, CH-8053 Z\"urich, Switzerland}
\date{March 6, 2005}
\maketitle

\begin{abstract}
The Poincar\'e mass operator can be represented in terms of a $Cl(3,0)$ Clifford algebra.
With this representation the quadratic Dirac equation and the Maxwell equations
can be derived from the same mathematical structure.
\end{abstract}

{\scriptsize\PACS{12.20.-m; 03.65.Fd; 03.50.De; 02.10.Hh; 02.10.De}
\MSC{81V10; 11E88; 30G35; 15A33; 81R20}
\UK{Hyperbolic complex Clifford algebra; Hyperbolic numbers; Mass operator; Quadratic Dirac operator; Maxwell equations}}
\section{Introduction}
The complex numbers are naturally related to rotations and dilatations
in the plane, whereas the so-called hyperbolic numbers can be
related to Lorentz transformations and dilatations in the two-dimensional Minkowski
space-time. The hyperbolic numbers are also known as perplex \cite{Fje86}, unipodal \cite{Hes91}, duplex \cite{Kel94} or split-complex numbers.

The hyperbolic numbers are the universal one-dimensional Clifford algebra \cite{Kel94}.
They have been applied by Reany \cite{Rea93} to 
2nd-order linear differential equations. A function theory for hyperbolic numbers
has been presented by Motter and Rosa \cite{Mot98}. Extensions to an n-dimensional
space have been given by several authors \cite{Rea94,Fje98,Fje298,Wum02}.

It was shown by Hucks \cite{Huc93} that a four-component Dirac spinor is equivalent 
to a two-component hyperbolic complex spinor, and that the Lorentz group is equivalent to a 
generalized SU(2). In this work Hucks also found that the operations of C, P, and T on 
Dirac spinors are closely related to the three types of complex conjugation 
that exist when both hyperbolic and ordinary imaginary 
units are present. Xuegang et al. investigated in this context 
the Dirac wave equation, Clifford algebraic spinors, a hyperbolic Hilbert space,
and the hyperbolic spherical harmonics in hyperbolic spherical polar coordinates 
\cite{Xue00,Xue200,Xue01}.

It has been shown by Baylis and Jones \cite{Bay89} that a $Cl(3,0)$ Clifford algebra
has enough structure to describe relativity as well as
the more usual $Cl(1,3)$ or $Cl(3,1)$.
Baylis represents relativistic space-time points as paravectors. He applied 
these paravectors to Electrodynamics \cite{Bay99}
(for another approach to relativity see Hestenes \cite{Hes03}).
The paravectors are also used in this work, where hyberbolic numbers are
taken to explicitly represent the $Cl(3,0)$ Clifford algebra. Therefore, this algebra 
is denoted here as hyperbolic algebra.

The Poincar\'e mass operator is arising in a natural way as an operator which
is invariant under rotations
and translations in four-dimensional space-time.
In the following it is shown that the Poincar\'e mass operator can be represented 
with the $Cl(3,0)$ hyperbolic algebra and it is then possible
to derive the quadratic Dirac equation and the Maxwell equations directly from this operator.

\section{Hyperbolic algebra}
\label{not}
In this work the 
hyperbolic numbers are used for the algebraic representation of relativistic vectors.
The hyperbolic numbers $z\in\bfa{H}$ are defined as  
\be
\label{beg}
z=x+jy\;,\hspace{2cm}x,\,y \in\bfa{C}\;,
\ee
where the hyperbolic unit $j$ has the property
\be
j^2=1\;.
\ee
In addition to the complex conjugation, 
a hyperbolic conjugation can be defined
which changes the sign of the
hyperbolic unit
\be
\label{conj}
\bar{z}=x-jy\;.
\ee 
The hyperbolic numbers are a commutative extension of the complex numbers to
include new roots of the polynomial equation $z^2 - 1 = 0$.
In the terminology of Clifford algebras they are represented by $Cl(1,0)$,
i.e. they correspond to a one-dimensional Clifford algebra. 

Using the hyperbolic unit 
an algebra for the description of relativistic vectors can be introduced.
A contravariant Lorentz vector with the
coordinates $p^\mu=(p^0,p^i)$
is represented as 
\be
\label{veco}
P=p^\mu\sigma_\mu\;.
\ee
The basis vectors $\sigma_\mu$ include the
unity and the elements of the Pauli algebra
multiplied by the hyperbolic
unit $\sigma_\mu=(1,j\sigma_i)$. Note that this algebra is congruent to the
$Cl(3,0)$ paravector algebra of Baylis \cite{Bay89,Bay99}.

\section{Poincar\'e mass operator}
\label{vec}
With the above vector representation the Poincar\'e mass operator can be introduced as a product of a momentum vector and
its hyperbolic conjugated counterpart
\be
\label{pmassop}
M^2=P\bar{P}\;.
\ee
$\bar{P}$ is a realization of the so-called Clifford conjugation, 
the involution of spatial reversal \cite{Bay99}.

The explicit form of the mass operator is obtained by a multiplication of the basis matrices.
The mass operator can be separated into a spin dependent and a
spin independent contribution  
\be
\label{spindef}
P\bar{P}=p_\mu p^\mu-i\sigma_{\mu\nu}p^\mu p^\nu\;,
\ee
where the spin operator is given by
\be
\label{sigspin}
\sigma_{\mu\nu}=
\left(\begin{array}{cccc}
\;0\;&\;-ij\sigma_1\;&\;-ij\sigma_2\;&\;-ij\sigma_3\;\\
\;ij\sigma_1\;&\;0\;&\;\sigma_3\;&\;-\sigma_2\;\\
\;ij\sigma_2\;&\;-\sigma_3\;&\;0\;&\;\sigma_1\;\\
\;ij\sigma_3\;&\;\sigma_2\;&\;-\sigma_1\;&\;0\;\\
\end{array}\right)\;.
\ee
Since the spin operator is anti-symmetric, the last term in Eq.~(\ref{spindef}) is in this case zero. However, the spin term becomes
inportant when interactions are considered.

A basic fermion equation can be introduced as an eigenvalue equation of the mass operator.
With the hyperbolic algebra defined above the
equation can be written as
\be
\label{equat}
M^2\psi(x)=m^2\psi(x)\;,
\ee
The wave function $\psi(x)$ has the general structure 
\be
\label{Ansatz}
\psi(x)=\varphi(x)+j\chi(x)\;,
\ee
where $\varphi(x)$ and $\chi(x)$ can be represented as two-component
spinor functions \cite{Huc93}. They depend on the four space-time
coordinates $x^\mu$. Equation~(\ref{equat}) is a 
second order differential equation 
in coordinate space. Therefore, the momentum vectors included in $M^2$ are replaced by the operators $p^\mu=i\partial^\mu$. 
 
\section{Quadratic Dirac equation}
\label{blab1}
The following considerations focus on the description of electrons and photons.
Electromagnetic interactions can be introduced with the minimal substitution of the
momentum operator. The mass operator of Eq.~(\ref{pmassop}) transforms into
\be
\label{basic}
M^2=(P-eA(x))(\bar{P}-e\bar{A}(x))\;,
\ee
and is now invariant under local gauge transformations.
This mass operator can be inserted into Eq.~(\ref{equat}). If Pauli matrices 
and electromagnetic fields are expressed with the
anti-symmetric tensor $\sigma_{\mu\nu}$ given in Eq.~(\ref{sigspin})
and $F^{\mu\nu}=\partial^\mu A^\nu- \partial^\nu A^\mu$ one finds
\be
\label{Pauli1}
\left((p-eA)_\mu(p-eA)^\mu -\frac{e}{2}
\sigma_{\mu\nu} F^{\mu\nu}-m^2\right)\psi(x)=\,0\;.
\ee

This is the quadratic Dirac equation. Using the hyperbolic algebra
it can be represented as a $2\times 2$ matrix equation, whereas conventionally
the quadratic Dirac equation is a $4\times 4$ matrix equation.
One finds in both cases the same two coupled
differential equations. In the hyperbolic formalism
the terms proportional to the hyperbolic unit include the
differential equation of the lower component, the other terms describe the
differential equation of the upper component. Conventionally,
the coupled differential equations for upper and lower components are separated by
the matrix structure.

\section{Maxwell equations}
\label{electro}
The Maxwell equations can be derived from an eigenvalue equation of the mass operator,
where the mass operator is now acting on a vector field
\be
\label{wavb}
M^2 A(x)=0\;.
\ee
The equation can be expressed
with the electromagnetic fields according to
\bea
\label{maxwell}
P\bar{P}A&=&-\bfa{\nabla}\cdot\bfa{E}-\partial^0 C\nonumber\\
                  &&+ij\bfa{\nabla}\cdot\bfa{B}\\
                  &&-j(\bfa{\nabla}\times\bfa{B}-
                       \partial^0\bfa{E}
                      -\bfa{\nabla}C)\nonumber\\
                  &&-i(\bfa{\nabla}\times\bfa{E}+\partial^0\bfa{B})
                   =0\nn\;.
\eea
This expression is obtained if one evaluates $\bar{P}A(x)$, inserts the usual definitions for the
electromagnetic fields, and then multiplies the resulting terms by the operator
$P$. If $P\bar{P}$ is calculated first, Eq.~(\ref{wavb}) reduces to the wave operator acting on the
vector potential giving zero. Both forms are equivalent in the Lorentz gauge. Note, that the Pauli algebra is implicitly 
part of the 3-dimensional vectors, e.g. $\bfa{E}(x)=E^i(x)\sigma_i$.

In Eq.~(\ref{maxwell}) the four homogeneous 
Maxwell equations are included. 
The calculation provides two additional terms depending on 
\be
\label{zero}
C(x)=\partial_\mu\, A^\mu(x)\;.
\ee
These terms disappear in the Lorentz gauge.

\section{Summary}
The representation of the Poincar\'e mass operator in terms of a $Cl(3,0)$
Clifford algebra can be used to construct equations of motion of relativistic quantum fields.
In general the representation has the form
\be
M^2= P\bar{P}\;,
\ee
where $P$ is the $Cl(3,0)$ momentum vector and 
$\bar{P}$ denotes the Clifford conjugation of $P$.
If more complex symmetry transformations like local gauge transformations are included,
the mass operator is modified (see Eq.~(\ref{basic})). The mass operator is then
a Casimir operator of the extended symmetry group, i.e. \hspace{-0.05cm}it is invariant also under the
additional symmetry transformations.

\end{document}